\documentstyle[preprint,aps]{revtex}
\begin{document}
\draft
\preprint{}
\title{CONFINEMENT, QUARK MATTER EQUATION OF
STATE AND HYBRID STARS}
\author{S.B. Khadkikar, A. Mishra, H. Mishra}
\address{Theory Group, Physical Research Laboratory, Navrangpura, Ahmedabad
380 009, India}
\maketitle
\def\be{\begin{equation}}
\def\ee{\end{equation}}
\begin{abstract}
We consider here quark matter equation of state in a relativistic
harmonic confinement model at zero temperature. The same is considered to
study phase transition of neutron matter to quark matter at
high densities. This, along with a phenomenological equation
of state in the neutron matter sector is used to study hybrid
stars. Using Tolman Oppenheimer Volkoff equations,
stable solutions for hybrid stars are obtained with a maximum mass
of 1.98 $M_\odot$ and radii around 10 kms with a quark core
of about 1 to 2 kilometers.
\vspace{2in}
\center{\bf (To appear in Mod. Phys. Lett. A)}
\end{abstract}
\pacs{}
\narrowtext
\section{INTRODUCTION}
It is believed that hadronic matter undergoes a phase transition
to quark matter at high temperatures and/or high densities.
 We may have these transitions possible in heavy ion collision experiments.
However, signals here may get masked since such a phase
lasts for a very short time
followed by hadronisation. On the other hand we could also have
compact stellar objects formed in a gravitational collapse where
quark gluon matter could exist at sufficiently high densities.
Since quarks and gluons are confined with absence of totally asymptotic
states the overall confinement can affect  the equation of
state for quark matter. A phase transition of neutron matter to quark matter
at zero temperature or temperature small compared to
degeneracy temperature allows the existence of hybrid stars i.e.,
stars having a quark core and crust of neutron matter with
appropriate pressure balancing at the
interface [1-6].
 In fact,
the large magnetic field of the pulsars could be held by quark matter
with nonzero electrically charged constituents rather than neutron matter.

Although there is a naive wish that quantum chromodynamics (QCD) at high
temperature and/or densities should approach free field limits, there have been
many rather convincing results that QCD remains as nonperturbative as
the confined phase at high temperatures \cite{negle}.
Expecting a similar behaviour at finite baryon densities also, we are
motivated to study thermodynamic properties, e.g.,
the equation of state at finite densities
including the effect of confinement explicitly.

In the present paper we first analyse the quark matter equation of state
that explicitly takes into account the overall confinement of the quarks
using a relativistic harmonic oscillator potential.
The harmonic confinement model as used here has been applied earlier
for study of hadronic properties
as well as to consider equation of state at
finite temperature and zero baryon density
\cite{jcpkha}. We consider this model
at finite baryon density as may be relevant for hybrid stars. In the neutron
matter sector
we shall use a phenomenological equation of state. This is then used to
consider phase transition to quark matter and to study hybrid stars.

It is generally felt that it is difficult to take correct thermodynamic
limit in the presence of confinement.  However the oscillator states of
the relativistic harmonic confinement model are exactly analogous to
the Landau orbits of electrons in a magnetic field \cite{jcpkha}.
The Landau orbits are characterised by an arbitrary centre of confinement.
The collection of such orbits has no difficulty in forming a large
system with appropriate thermodynamic limit. However, if few such orbits
form a colourless bound state like a hadron the centre of mass of the
system is free to move and the centre of mass correction of the
bound state has to be applied. In the context of neutron stars
the gravitational field crunches the neutrons and a large macroscopic
quark matter consisting of the Landau like states is formed.
In any confinement model the energy
per particle grows rapidly with the number of particles.This would mean
that one requires infinite energy in the thermodynamic
limit. However, one has to keep in mind that the gravitational
potential which grows as square of the number
of particles is primarily responsible in producing the
macroscopic state.
The thermodynamic properties of this system are calculated by using the
partition function which does not involve gravitational field but
takes account of modified phase space occupation because of long range
confinement. Then the full system is considered inclusive of
gravitational field in TOV equations. The analogue of an electron in a
magnetic field is used for obtaining the statistical weights in
converting momentum space integrals to sums over possible states in the
confining field.

We organise the paper as follows. In section 2, we derive the equation
of state for quark matter in relativistic harmonic confinement
model and use Gibbs criterion to consider the possibility
of a phase transition of neutron matter to quark matter. In section 3 we
shall use these equations of state to consider hybrid stars
using Tolman Oppenheimer Volkoff equations. In section 4, we discuss and
summarise our results.

\section{Relativistic harmonic confinement model and quark matter
equation of state}

In the harmonic confinement model \cite{khagup} the relativistic
equation for the quarks is given as
\be
[i\gamma^\mu \partial _\mu -M_i-V(r)]\psi_i(\vec r)=0
\ee
where $V(r)$ is the `potential' with a Lorentz scalar plus
vector structure given as
\be
V(r)=\frac{1}{2}(1+\gamma_0)\alpha^2 r^2.
\ee
In the above, the subscript `i' in $\psi(r)$ represents the
colour and flavour index of the quark.
This model had been applied earlier to study hadronic properties
\cite{khagup} and nucleon nucleon angular distributions \cite{3viju}.
In terms of the two component single particle wave functions,
equation (1) may be written in the simple form
\be
[p^2+\Omega ^2 r^2]\phi_i(\vec r)
=(E^2-M_i^2)\phi_i(\vec r),
\ee
where $\Omega =\alpha \sqrt{E+M_i}$ are $"$frequencies"
(assumed to be constant) of the quark fields. We may note that the
confining term $\Omega^2 r^2$ is given by $Tr(A_\mu^bA^{\mu b})$,
where $A_\mu^b$ is the background field \cite{jcpkha} similar
to quantum motion of an electron in external magnetic field giving
rise to Landau orbits. The `potential'
$V(r)$ in equation (2) which has both Lorentz scalar and
vector components gives rise to the energy levels as
\cite{jcpkha,khagup}
\be
{E_n^i}^2=M_i^2+(2n+1)\Omega \quad (n=1,2,\cdots).
\ee
For the study of hadronic properties the parameters chosen
were \cite{khagup}
$$M_{u,d}=160.66 MeV$$
$$M_s=460.66 MeV$$
$$\sqrt \Omega=1.162 fm^{-1},$$
which we shall be using to derive the quark matter equation
of state. The thermodynamic pressure $P_i$ (i=u,d,s) is then given by
\cite{huang}
\be
\frac{P_i V}{k_B T}=g_Vg_I\sum_n g_n \left[
log\Big(1+e^{-\beta(E_n^i -\mu^i)}\Big)
+log\Big(1+e^{-\beta(E_n^i +\mu^i)}\Big)\right],
\ee
where $g_V$, $g_I$ and $g_n$ are respectively the statistical
weights corresponding to volume, internal degrees of freedom
(spin and colour) and the degenaracy factor  for the nth energy
level. These are given as
 $g_V=4\pi/3\Omega^{3/2}\times V/(2\pi)^3$ \cite{com},
$g_I=2\times 3$ and $g_n=n(n+1)/2$
\cite{jcpkha,huang}. Further, $\mu^i$ is the chemical potential
associated with the $i$-th quark. We may note that no correct
thermodynamic limit can be considered when there is harmonic confinement
as the system does not have translational invariance. One has to include
center of mass motion corrections to be able to obtain correct
thermodynamic limit. Such corrections, though important for hadron
spectroscopy [9] and nucleon - nucleon scattering [10] involving few
particles, are expected to be small ($\simeq$ 1/N, N is the number of
particles) and are not included.

As we shall be considering hybrid stars, we shall consider the
equation of state at zero temperature and for quark matter,
along with u and d quarks, we include strange quarks and electrons
which could be produced at high densities through weak
interactions and be in equilibrium. The chemical potentials for
each species of fermion may be written down in terms of the two
independent chemical potentials $\mu_B$ and $\mu_E$, the baryon
and electric charge chemical potentials respectively as
$$\mu_u=\frac{1}{3}\mu_B+\frac{2}{3}\mu_E
\label{mua}\eqno(6a)$$
$$\mu_d=\frac{1}{3}\mu_B-\frac{1}{3}\mu_E
\label{mub}\eqno(6b)$$
$$\mu_s=\frac{1}{3}\mu_B-\frac{1}{3}\mu_E
\label{muc}\eqno(6c)$$
$$\mu_e=-\mu_E
\label{mud}\eqno(6d)$$
\setcounter{equation}{6}

Thermodynamic pressure for quark matter is then given as
\be
P_{qm}=\sum_{i=u,d,s,e}P_i
\ee
where the electron pressure is given as
\be
P_e=\frac{\mu_e^4}{12\pi^2}
\ee
These chemical potentials are fixed from the electrical charge
 neutrality condition for a given baryon density.
The baryon number density and the
electric charge densities are given as
\be
\rho_B=\frac{1}{3}\sum_{i=u,d,s}\rho_i
\ee
and
\be
\rho_E=\frac{2}{3}\rho_u-\frac{1}{3}\rho_d-\frac{1}{3}\rho_s
-\rho_e,
\ee
where the quark densities at zero temperature are
\be
\rho_i=\sum_n \theta(\mu^i-E_n^i)g_Vg_Ig_n/V
\ee
and
the electron density is
\be
\rho_e=\frac{\mu_e^3}{3\pi^2}.
\ee
It is then straightforward to evaluate the sums in equation (5)
for each quark flavour to obtain the equation of state.

In the neutron matter sector, the equation of state is still a
matter of debate particularly at higher densities. Different
 parametrisations for the same are there for the description
of the structure of neutron stars \cite{arnett}.
In contrast to sophisticated parametrisations designed
for neutron star matter, we shall use throughout in this
paper a simple parametrisation for the neutron matter equation
of state by Sierk and Nix \cite{sierk}
\be
\epsilon_N (\rho)=\rho\left[\frac{2}{9}K\Big(
\sqrt{\frac{\rho}{\rho_0}}-1\Big)^2+W_0
+W_{sym}+M_N\right].
\ee
In the above, $W_0=-16$ MeV is the  binding energy per nucleon at normal
nuclear matter density, $\rho_0$=0.145 fm$^{-3}$, $M_N$=939 MeV,
the rest mass of neutron and $W_{sym}$=32 MeV is the symmetry energy
of neutron matter as estimated from liquid drop model calculations.
While deriving the equation of state for the neutron matter,
we take the value of compressibility, K=550 MeV \cite{rosen}.
Below the nuclear matter densities the star matter is no longer determined
by nucleon nucleon interactions only and we shall use for subnuclear
density regions the equation of state given by Baym, Bethe and
Pethick \cite{bbp} in the region 0.00267
 fm$^{-3}<\rho<$1.6fm$^{-3}$ and
the equation of state given by Baym, Pethick and Sutherland
\cite{bps} for densities less than the above region.
Different density regions of each equation of state were
so taken that the pressure and energy density varied smoothly
from one equation of state to the other \cite{rosen,shapiro}
in the neutron matter sector.

Once the energy density is known the pressure at zero temperature
can be obtained from the thermodynamic relation \cite{fetter}
\be
P_{nm}(\rho)=\rho\frac{\partial \epsilon_N}{\partial \rho}-\epsilon_N,
\ee
and the chemical potential $\mu_B$ is given as
\be
\mu_B=\frac{\partial \epsilon_N}{\partial \rho}.
\ee

We shall now consider the scenario of phase transition from cold
neutron matter to quark matter. As usual, the phase boundary
of the coexistance region will be given by Gibbs criteria.
The critical pressure and the critical chemical potentials
are given by the condition
\be
P_{nm}(\mu_B)=P_{qm}(\mu_B).
\ee
The left hand side of the above equation is given through equations
(13) and (14) and the right hand side is
 the zero temperature limit of equations (5) and (7).

We may note that in the calculations of pressure and chemical potentials
in equations (5), (9) and (10) in the quark matter sector the zero
point energy and in equations (13), (14) and (15) for the neutron
matter sector the nucleon mass have been subtracted. This is
because we wish to compare the pressure and chemical potential
to consider phase transition for which the ground state energy is
not relevant. However these will be included in the energy densities
when we consider hybrid star as they contribute to the
gravitational effects.

In Fig. 1, we plot (P,$\mu_B$) curves for quark  matter
and neutron matter which yield the critical parameters
$P_{cr}=300 MeV/fm^3$ and $(\mu_B)_{cr}$=610 MeV.
In Fig. 2, we have plotted the baryon number density in
the quark matter sector as a function of the baryon chemical
potential. The shape of the curve is indicative of the discrete
energy levels characteristic of an oscillator potential.
The baryon number density in the quark matter sector
corresponding to the critical $\mu_B$ is calculated
to be $\rho_B^{qm}=1.341fm^{-3}$. The same in the neutron matter
sector at the critical $\mu_B$ is $\rho_B^{nm}$=0.784fm$^{-3}$,
about 5.4 times the nuclear matter density,
which seems reasonable.
At the critical pressure, the
energy densities for the quark matter and neutron matter
sectors, $\epsilon_{cr}^{qm}$ and $\epsilon_{cr}^{nm}$
are found to be 2542 MeV$/fm^3$ and 918 MeV$/fm^3$ respectively.
The discontinuity in the
number density as well as energy density indicates a first order
phase transition. We may note here that the calculation
for the two phases are done in two different
models-a phenomenological parametrisation in the
neutron matter sector and the oscillator confinement model in the quark
matter sector. The different treatments of the two phases essentially
leads to a first order phase transition as above.
\section{Hybrid stars}
For the description of neutron star, which is highly concentrated
matter so that the metric of space-time geometry is curved, one has to
apply Einstein's general theory of relativity. The space-time
geometry of a spherical neutron star described by a metric which in
Schwarzschild coordinates has the form \cite{rosen,weber}
\begin{equation}
ds^2=-e^{\nu(r)}dt^2+[1-2M(r)/r]^{-1}dr^2+r^2[d\Theta^2+sin^2 \Theta
d\phi^2]
\end{equation}
The equations which determine the star structure and the geometry are,
in dimensionless forms,
\begin{equation}
{d\hat P(\hat r r_0)\over d\hat r}=-\hat G
{[\hat\epsilon (\hat r r_0)+\hat P (\hat r r_0)][\hat M (\hat r r_0)
+4\pi a \hat r^3 \hat P(\hat r r_0)]\over \hat r^2[1-2\hat G
\hat M (\hat r r_0)/\hat r]},\label{tov1}
\end{equation}
\begin{equation}
\hat M (\hat r r_0)=4\pi a \int_0^{\hat r r_0} d\hat r^\prime
\hat r^{\prime^2} \hat \epsilon(\hat r^\prime r_0),\label{tov2}
\end{equation}
and the metric function, $\nu (r)$, relating the element of time at
$r=\infty$ is given by
\begin{equation}
{d\nu(\hat r r_0)\over d\hat r}=2\hat G {[\hat M (\hat r r_0)
+4\pi a \hat r^3 \hat P(\hat r r_0)]\over \hat r^2[1-2\hat G
\hat M (\hat r r_0)/\hat r]}.\label{tov3}
\end{equation}
In equations (18) to (20) the following substitutions have been made.
\begin{equation}
\hat \epsilon\equiv \epsilon/\epsilon_c,\quad
\hat P\equiv P/\epsilon_c,\quad\hat r\equiv r/r_0,\quad
\hat M\equiv M/M_\odot,
\label{tov4}
\ee
where, with
$ f_1=197.329 \; MeV\; fm$ and $r_0=3\times 10^{19}$ fm,
we have
\begin{equation}
a\equiv\epsilon_c r_0^3/M_\odot, \quad
\hat G\equiv (G/f_1)/(r_0/M_\odot)
\label{tov5}
\end{equation}
In the above, quantities with hats are dimentionless.
G in equation (\ref{tov5}) denotes the gravitational constant
$(G=6.707934\times 10^{-45}\;\mbox{MeV}^{-2})$.

Further the surface condition for the metric function is given by
\[ \nu(R)=ln(1-2GM/R) \]

In order to construct a stellar model, one has to integrate equations
(\ref{tov1}) to  (\ref{tov3}) from the star's center at r=0 with
a given central energy density $\epsilon_c$ as input until the
pressure $P(r)$ at the surface vanishes. As stated, with
central density greater than the critical energy density
as dictated by the last section, we expect that we
shall have quark matter, and not neutron matter at the center of the
star. Hence we shall be using here
the equation of state for quark matter using
the zero temperature limit of equation (7) with
$\hat P(0)=P(\epsilon_c)$. We then integrate the TOV equations
until the pressure
and density decrease to their critical values, so that there is a first
order phase transition from quark matter to neutron matter at radius
$r=r_c$ for a given central density. For $r>r_c$, we shall have equation
of state for neutron matter where  pressure will change continuously
(but the energy density will have a discontinuity at $r=r_c$).
The TOV equations with neutron matter equation of state
as given in equations (13) and (14) are continued.
As described earlier we
use the equation of state by Baym, Bethe and Pethick \cite{bbp}
for subnuclear density region of neutron rich nuclei
and of Baym, Pethick and Sutherland \cite{bps} for the lower
density crystalline lattice bathed in relativistic electrons.
The TOV equations are continued with the above equations of
state till the pressure vanishes which defines the surface of
the star. This will complete the calculations for stellar
model for a hybrid ``neutron" star,
whose mass and radius can be calculated for different central densities.

The energy density profile obtained from (\ref{tov1}) to (\ref{tov3}) are
plotted in Fig. 3 for central densities $\epsilon_c=3250$ MeV/fm$^3$
and $\epsilon_c=4000$ MeV/fm$^3$.
For core energy densities greater than the critical energy
density ($\simeq 2542$ MeV/fm$^3$) the core consists of quark
matter. As we go away from the core towards the surface through TOV
equations, when the critical pressure is reached, the density
drops discontiniously indicating a first order phase transition.
 Thus e.g. for central density of 3250 MeV/fm$^3$ such a star has a quark
matter core of radius 1 km with a crust of neutron matter of about 10 kms,
whereas for $\epsilon_c$=4000 MeV/$fm^3$, the quark matter core radius is
1.8kms with neutron matter crust of 9 kms.

Fig. 4 depicts mass of a hybrid star as a function of its
central energy density. Here we shall have two branches of solutions.
Pure neutron matter stars at lower central densities
$\epsilon_c <\epsilon _{nm}^{cr}$
and hybrid stars at central
energy densities  $\epsilon_c >\epsilon _{qm}^{cr}$.
Taking into account the stability of such stars under
density fluctuations require $dM/d\epsilon_c>0$ \cite{weinberg}.
We find stable hybrid stars with central energy densities
$\epsilon_c \approx 3-4.6 GeV/fm^3$ and radii around
10 to 11 kilometers, the critical mass being around
1.98 $M_\odot$, beyond which they may collapse to
black holes. The instability region is
shown as dashed curves in the above figure.

Fig. 5 describes the mass radius relationship for hybrid star.
This is important to estimate how rapidly the stars with a given
equation of state can rotate without mass shedding at the
equator.

We may take the relativistic Keplerian angular velocity $\Omega_K$ given as
\cite{friedman}
\be
\frac{\Omega_K}{10^4 sec^{-1}}=0.72\sqrt{\frac{M/M_\odot}{(R/10km)^3}}.
\ee
as for neutron stars and estimate the same. Figure 6 shows the variation of
$\Omega_k$ as a function of its mass for such a star.

We wish to estimate the moment of inertia of these stars. The
expression for the moment of inertia of pulsars as well as the
equillibrium for the rotating hybrid stars is given
as \cite{hmpkp,hartle,glewmos}
\be
I=\frac{8\pi}{3}\epsilon_c r_0^5\int_0^{\hat R r_0}
d\hat r {\hat r}^4\frac{[\hat\epsilon(\hat r r_0)
+\hat P(\hat r r_0)]
\exp(-\nu(\hat r r_0)/2)}{\sqrt{[1-2\hat G\hat M(\hat r r_0)/\hat r]}}.
\ee

We further estimate the surface gravitational red shift $Z_s$ of
photons as \cite{brecher}
\be
Z_s=\frac{1}{\sqrt{[1-2G\;M/R]}}-1.
\ee
This is plotted in Figure 7 as a function of mass.
In this context it may be worth while to mention here
that the surface redshifts as determined from gamma ray bursters
seem to lie in the range 0.2 to 0.5 \cite{liang} where as
the masses seem to lie in the range 1 to 1.85 $M_{\odot}$
with an error of $\pm 0.3$ at both ends \cite{joss}.

\section{Conclusions}
Let us summarise the findings of the present paper.

Using a relativistic harmonic confinement model for
quark matter \cite{khagup} and with a phenomenological
parametrisation for neutron matter \cite{sierk}
we saw that a first order phase transition exists between
the neutron phase and quark phase
at about five to six times the nuclear matter density.

We have used here the paramters of the harmonic confinement
model fixed from hadronic properties.
In this context or otherwise, we may note that the harmonic
confinement model has its origin through Prasad -- Sommerfeld
dyon like configurations in QCD \cite{jcpkha,khagup}.
When we include temperature and/or density effects we
may expect that they will affect such configurations in a
dynamical manner \cite{hmpkp,hmam} and the parametrs of the
confinement model may change. However to get an
insight into the equation of state including the effect of
harmonic confinement we have approximated $\Omega$ to be independant of
temperature and/or density.

To study the possibility of a hybrid star, namely a star consisting of both
quark matter and neutron matter, we applied the TOV equation to the appropriate
equation of state with a given central energy density $\epsilon_c$.
It turns out that a stable hybrid star with a quark core and a neutron
matter crust can exist upto $\epsilon_c\simeq 4.6$ GeV/fm$^3$ beyond
which instability may result. For $\epsilon_c$ from 3000 to 4600 MeV/fm$^3$,
the mass of the star varies between 1.795 to 1.98 $M_\odot$ and the radius
of the star between 10 to 11 kms.
Consistently, the bulk of the hybrid star is provided by the neutron matter,
the quark matter providing  core of about 1 to 2 kms. The detailed
properties of neutron matter e.g. whether it is $soft$ or $stiff$ do not
seem to be reflected in the gross properties of hybrid stars.

\acknowledgements
The authors are thankful to A.R. Prasanna, J.C. Parikh for
discussions. AM and HM
acknowledge many discussions with S.P. Misra, N. Barik and P.K. Panda.

\newpage
{\bf {\Large Figure captions}}

\bigskip
\noindent {\bf Fig.1:} We plot pressure, P in MeV/fm$^3$
as a function of the baryon chemical potential, $\mu_B$
in MeV for the quark matter (the solid line)
and neutron matter (dashed line).
\hfil
\medskip

\noindent {\bf Fig.2:} We plot here
baryon number density, $\rho_B$ in fm$^{-3}$
as a function of baryon chemical potential $\mu_B$ in MeV for
quark matter.
\hfil
\medskip

%
\noindent {\bf Fig.3:} We plot  here the energy profile curves
inside the hybrid star for central densities 3250 MeV/fm$^3$
(the solid line) and 4000 MeV/fm$^3$ (dot dashed line).
Discontinuity at the critical energy density is shown by
the dashed line.
\hfil
\medskip

\noindent {\bf Fig.4:} Mass of the hybrid star as a function
of central density, $\epsilon_c$ in MeV/fm$^3$
 is plotted here. The dashed portion
indicates instability.
\hfil
\medskip

\noindent {\bf Fig.5:} We plot here mass of the star in units of solar mass
as a function of radius of the star in kilometers.
\hfil
\medskip

\noindent {\bf Fig.6:} We plot  here the Keplerian angular velocity,
$\Omega_K$ in 10$^4sec^{-1}$
as a function of $M/M_\odot$.
\hfil
\medskip

\noindent {\bf Fig.7:} We plot  here the surface gravitational
red shift, $Z_s$ as a function of $M/M_\odot$.
\hfil
\medskip

\begin{thebibliography}{99}
\bibitem{com} There is a difference of factor $4\pi/3$ as
compared to Ref. 7 in the expression of $g_V$. When this factor is
taken into account, the critical temperature of Ref. 7
reduces from 170 MeV to 110 MeV.
\end{thebibliography}
\end{document}